\documentclass[Afour,sagev,times]{sagej}

\usepackage{moreverb,url,subfigure}

\usepackage[colorlinks,bookmarksopen,bookmarksnumbered,citecolor=red,urlcolor=red]{hyperref}

\newcommand\BibTeX{{\rmfamily B\kern-.05em \textsc{i\kern-.025em b}\kern-.08em
T\kern-.1667em\lower.7ex\hbox{E}\kern-.125emX}}

\def\volumeyear{2021}

\begin{document}

\runninghead{Fern\'{a}ndez-Guillam\'{o}n and Molina-Garc\'{i}a}

\title{Simulation of Variable Speed Wind Turbines based on Open-Source Solutions: application to Bachelor and Master Degrees}

\author{Ana Fern\'{a}ndez-Guillam\'{o}n\affilnum{1} and \'{A}ngel Molina-Garc\'{i}a\affilnum{1}}

\affiliation{\affilnum{1}Dept. of Automatics, Electrical Engineering and Electronic Technology, Universidad Politecnica de Cartagena, 30202 Cartagena, Spain.}

\corrauth{Ana Fern\'andez-Guillam\'on}

\email{ana.fernandez@upct.es}

\begin{abstract}
  \textcolor{black}{This paper describes variable speed wind turbine (Types 3 and 4, IEC
  61400-27-1) simulations based on an open-source solution to be applied
  to Bachelor and Master Degrees. It is an attempt to improve the
  education quality of such sustainable energy by giving an open-source
  experimental environment for both undergraduate and graduate
  students. Indeed, among the renewable sources, wind energy is
  currently becoming essential in most power systems. The
  simulations include both one--mass and two--mass mechanical models, as
  well as pitch angle control. A general overview of the structure,
  control, and operation of the variable speed wind turbine is provided by these
  easy-to-use interactive virtual experiments. In addition, a
  comparison between commercial and open-source software packages is
  described and discussed in detail. Examples and extensive results
  are also included in the paper. {\color{black}{The models are available in
  Scilab-Xcos file exchange for power system education and
  researcher communities.}}}
\end{abstract}


\maketitle

\section{Nomenclature}

\noindent 
\begin{tabular}{@{}ll}
	BSc & Bachelor degree\\
	DC & Direct current\\
	DFIG & Doubly fed induction generator\\
	ECTS & European Credit Transfer System\\
	EHEA & European Higher Education Area\\
	FOSS & Free open-source solution\\
	FSWT & Fixed speed wind turbine\\
	MSc & Master degree\\
	PhD & Philosophi$\ae$ doctor\\
	RES & Renewable energy sources\\
	TSO & Transmission system operator\\
	VSWT & Variable speed wind turbine\\
	WT & Wind turbine\\
\end{tabular}

\section{Introduction}

\begin{figure}[tbp]
	\centering
	\includegraphics[width=0.95\linewidth]{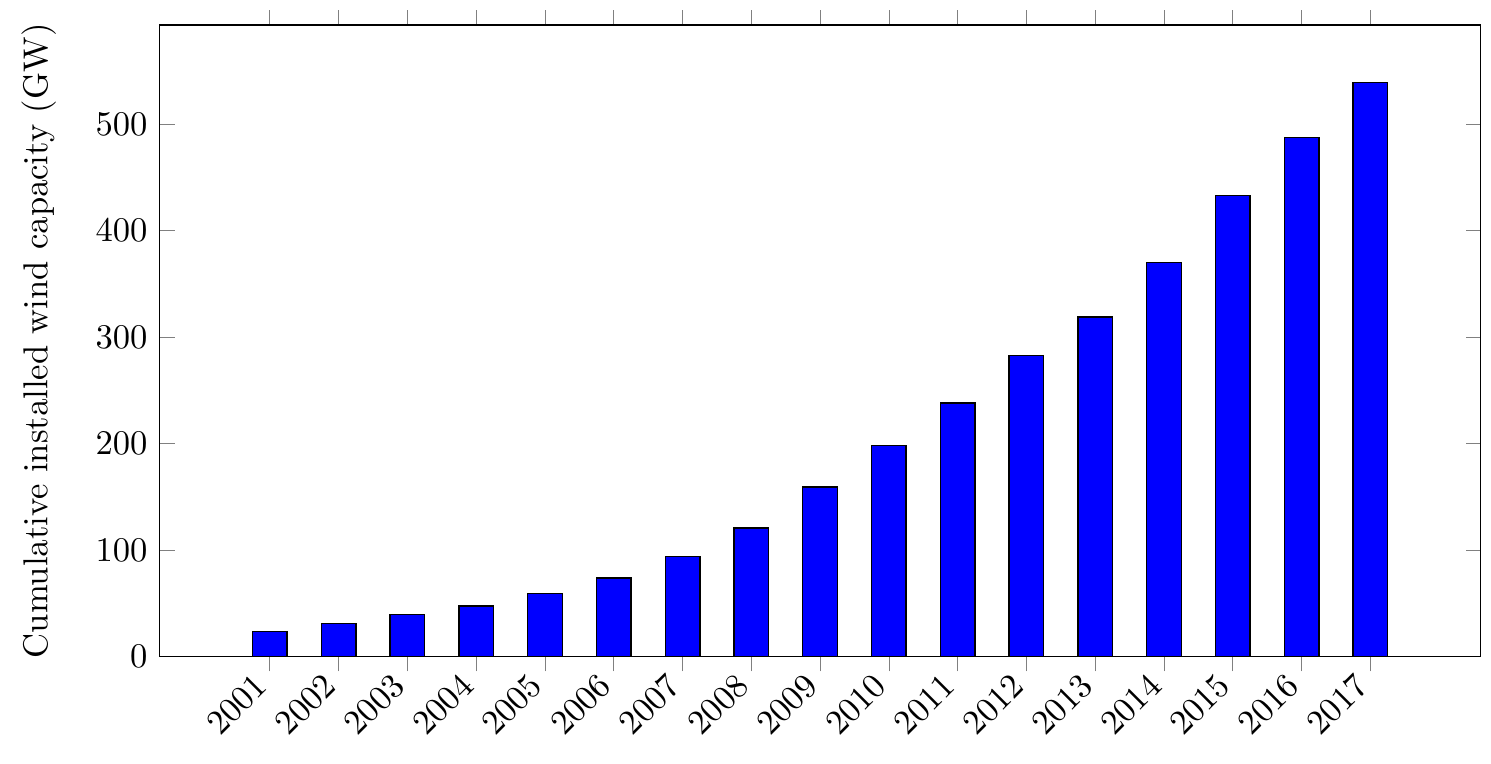}
	\caption{Global cumulative installed wind capacity in GW. Data from~\cite{gwec}}
	\label{fig.capacity}
\end{figure}

\begin{figure}[tbp]
	\centering
	\includegraphics[width=0.47\textwidth]{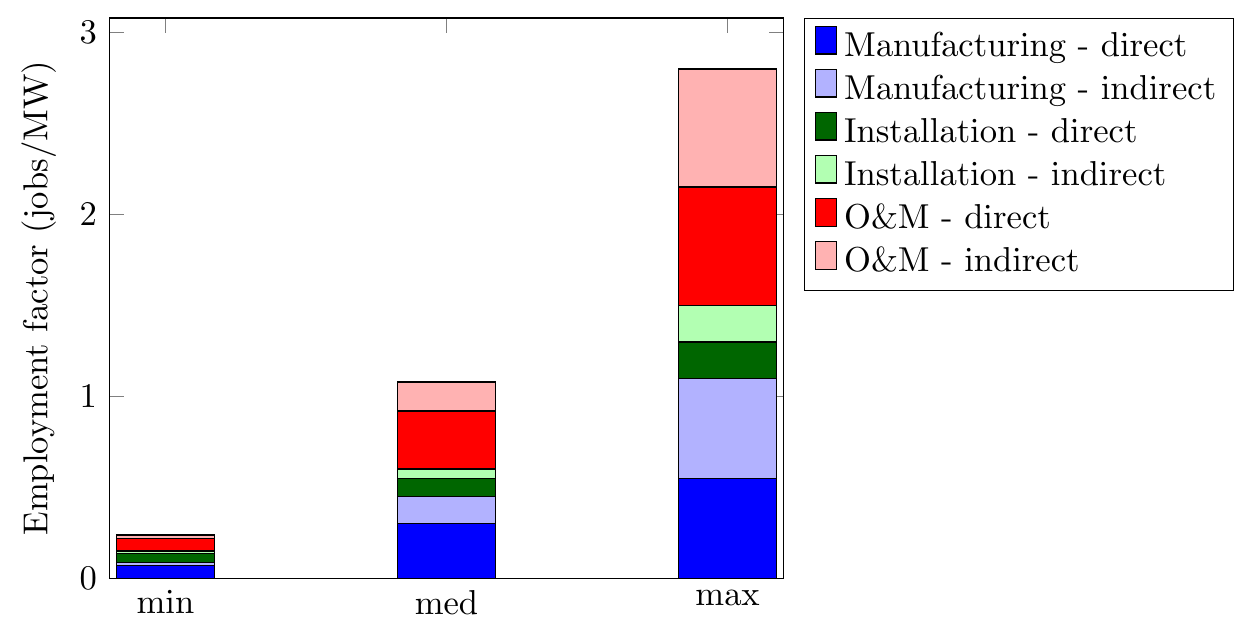}
	\caption{Direct and indirect jobs per deployment phase (in jobs/MW) for wind power. Data from \cite{cameron15}}
	\label{fig.jobs}
\end{figure}

\textcolor{black}{During the last decade, aspects such as climate change, energy dependence, fossil resource scarcity and the increasing costs of nuclear power, have promoted the integration of Renewable Energy Sources~(RES) into power systems~\cite{fernandez18}. Among these
renewable resources, wind power is the most popular alternative developed and currently integrated into the grid~\cite{fernandez2019offshore}. Actually, since 2001, the global
cumulative installed wind capacity has suffered an exponential growth,
see Figure~\ref{fig.capacity}. As a consequence, the wind energy
sector is promoting new employment opportunities~\cite{fragkos18}. 
These employment opportunities include direct employment ---manufacturing companies, promotion, utilities, engineering and R\&D--- as well as indirect employment ---providing services or components for wind turbines~(WTs)---~\cite{kaldellis11}.  Figure~\ref{fig.jobs} shows the normalised employment (jobs/MW) for manufacturing, installation, and O\&M, including direct and indirect employments determined by Cameron and Van Der Zwaan. 
However, and in spite of this relevant employment opportunity, it was
reported in 2017 a 'lack of talent' in the renewable energy
sector~\cite{swift19}. Consequently, it has been demanded by the sector new educational programs aiming to meet the increasing human capital
requirement for RES, especially in wind power areas~\cite{duran16}. Moreover, these activities require heterogeneous educational backgrounds, involving engineering, technician, economics, marketing, management and customer services~\cite{sooriyaarachchi15}. To overcome these deficiencies, some European universities have already started to implement new bachelor degrees~(BSc), master degrees~(MSc) and philosophi$\ae$ doctor~(PhD), as summarised in Table~\ref{tab.universities}.
Other universities include renewable energy subjects in industrial
engineering BSc, either as core academic or optional subjects.}

\begin{table}[tbp]
	\small\sf\centering
	\caption{European universities including new BSc, MSc and PhD related to energy and RES}
	\resizebox{\linewidth}{!}{%
		\begin{tabular}{llll}
			\toprule
			Degree & Name & University & Country  \\ 
			\midrule
			BSc & Energy engineering & University of Southern Denmark& Denmark\\
			BSc & Energy engineering & Aalborg University	& Denmark\\
			BSc & Energy engineering & Universidad Carlos III de Madrid	& Spain\\
			BSc & Energy engineering & Universidad Polit\'{e}cnica de Madrid	& Spain\\
			BSc & Energy engineering & Universidad Polit\'{e}cnica de Catalu\~{n}a	& Spain\\
			BSc & Renewable energy engineering & Stuttgart University	& Germany\\
			BSc & Renewable energy engineering & Berlin International College	& Germany\\
			BSc & Renewable energy engineering & Universidad Aut\'{o}noma de Barcelona	& Spain\\
			BSc & Renewable energy engineering & Universidad del Pa\'{i}s Vasco	& Spain\\
			BSc & Renewable thermal \& power engineering & National Research University& Russia\\
			& & & \\
			MSc & Renewable energy engineering & Carl Von Ossietzky Universität Oldenburg	& Germany \\
			MSc & Renewable energy engineering & Universidad Polit\'{e}cnica de Cartagena	& Spain\\
			MSc & Renewable energy engineering & University of Aberdeen	& UK \\
			MSc & Renewable energy engineering & Kingston University	& UK \\
			MSc & Renewable energy engineering & Heriot Watt University	& UK \\
			MSc & Wind energy & Technical University of Denmark	& Denmark\\
			MSc & Wind energy & Norwegian University of Science and Technology	& Norway\\
			MSc & Wind energy & Universidad Nacional de Educaci\'{o}n a Distancia			& Spain\\
			& & & \\
			PhD & Renewable energy & Aalborg University	& Denmark \\
			PhD & Renewable energy & Universidad Polit\'{e}cnica de Cartagena	& Spain \\
			PhD & Renewable energy & Universidad de Ja\'{e}n	& Spain \\
			PhD & Wind energy & Technical University of Denmark	& Denmark \\
			PhD & Wind energy & The University of Sheffield	& UK \\
			\bottomrule
		\end{tabular}
	}
	\label{tab.universities}
\end{table}

\textcolor{black}{Engineering students need to experiment and observe what they are taught in the theory sessions to improve their knowledge in topic~\cite{cedazo07}. In fact, virtual experimenting is considered an alternative solution for the practical sessions of the 'Wind energy' subject, facing the lack of wind energy experimental laboratories mainly due to the expensive costs of such teaching equipment. Laboratory
scaled-workbench solutions{\cite{5620949}} and lab-scale experiments {\cite{7528239}} have been previously proposed to overcome the high cost of such equipment. E-learning trends have also been considered by some authors as an alternative to lab-scale experiments. 
Other authors have developed a remarkable amount of modules for wind
teaching purposes mostly based on commercial software
packages~\cite{bentounsi10,tortoreli17}
(Matlab, Multisim, LabView, DigSilent) or packages that must run
within commercial solutions, such as SimPower {\cite{7339813,doi:10.1177/0309524X16642058}} or MatDyn {\cite{5598553}}. 
%
%
%
During the last decade, a continued effort to promote the use of Free Open-Source Solutions (FOSS) for engineering education
has been developed under different projects, taking into account the
educational and pedagogical benefits provided by such open-source
software {\cite{5275920,Hsu:2012:EFO:2417499.2417741}} 
Several advantages are found if comparing FOSS with commercial software, such as reducing costs of licenses, prevention of illegal copying and promoting self-learning and independent study~\cite{nehra14}}.

\textcolor{black}{Under this scenario, the present paper focuses on the simulation of Variable Speed Wind Turbines (VSWTs) within the 'Wind energy' subject by using the Scilab open-source solution. These practical sessions are currently included in both Electrical Engineering BSc and Renewable Energy MSc ---Universidad Polit\'{e}cnica de Cartagena, Spain---. Under this framework, we provide our students a visual, friendly and open-source tool for VSWT simulation purposes, including WT parameter analysis and a comparison between two WT mechanical models commonly used by researchers~\cite{seixas16}: one--mas and two-mass mechanical models. These simulations give the students a comprehensive view of VSWTs, and allow them to analyse their main curves. The contributions of the paper can be then summarised as:}
\begin{enumerate}
\item[(i)] Using an open-source mathematical software for simulation
  purposes.
\item[(ii)] \textcolor{black}{Analysing how wind speed variations can modify the pitch
  angle and, subsequently, the power coefficient and mechanical power of VSWTs.}
\item[(iii)] Determining the optimum tip speed ratio and power
  coefficient for different pitch angles, including the mechanical
  power and turbine rotational speed values.
\item[(iv)] Comparing the responses of generator and turbine
  rotational speeds depending on the mechanical model used.
\end{enumerate}
These objectives are developed under a teaching scenario according to
the current European educational framework. The models are available
in Scilab-Xcos file exchange for power system education and
researcher communities. 





\section{Educational framework in Europe: EHEA}\label{sec.educational_framework}

In 1999, 29 European Ministers of Education signed {\em{'The Bologna
    Declaration'}}, considered as the first attempt to create a new
European Higher Education Area (EHEA)~\cite{elias10}. Nowadays, this
EHEA involves 48 countries, 
and the main key-points can be summarised as follows: 
\begin{itemize}
\item Harmonisation of the Grade-system to promote European mobility
  students among the different countries. It involves the development
  of comparable criteria and methodologies among the universities, as
  well as the recognition of foreign degrees in other institutions and
  countries.
\item \textcolor{black}{Introduction of the European Credit Transfer System (ECTS). Each
  ECTS credit accounts for between 25-27~hours, including the learning
  hours of the students. Therefore, assuming that students devote
  40~h/week to study and learning tasks, each Grade level academic session should account for a maximum of 60~ECTS.}
\item Harmonisation of the teaching stages in all EHEA countries: BSc,
  MSc, and PhD. To access the following teaching cycle, it is
  required a successful finishing of the previous stage (BSc or MSc,
  respectively).
\end{itemize}
In most European countries, EHEA has thus promoted a significant
change in the universities, moving from the traditional
teacher-oriented approach to a more learner-centred
approach. 
Moreover, in West European countries ---such as Austria, Germany, and Spain---, the Bologna Process has been considered as a tool of domestic leverage by governments aiming to legitimate much wider projects of structural reform in the higher education sector {\cite{Harmsen2015}}.



\section{Wind turbines. Preliminaries}\label{sec.preliminaries}



\begin{figure}[tbp]
  \centering
  \includegraphics[width=0.4\textwidth]{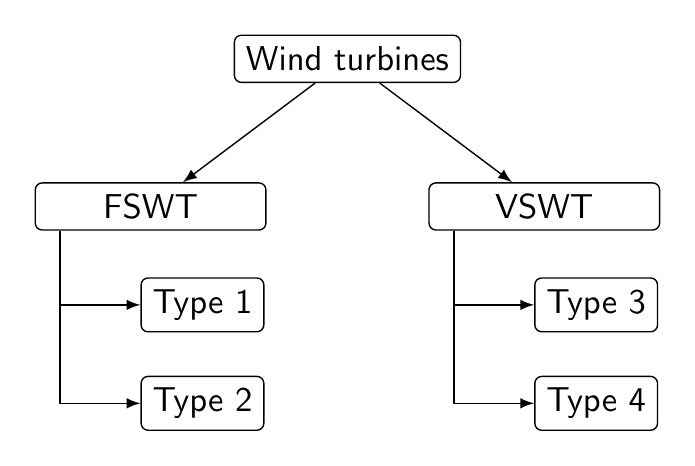}
  \caption{Classification of wind turbines.}
  \label{fig.clasification}
\end{figure}

\begin{figure}[tbp]
  \centering
  \subfigure[DFIG wind turbine. {\em{Type 3}}]{\includegraphics[width=0.35\textwidth]{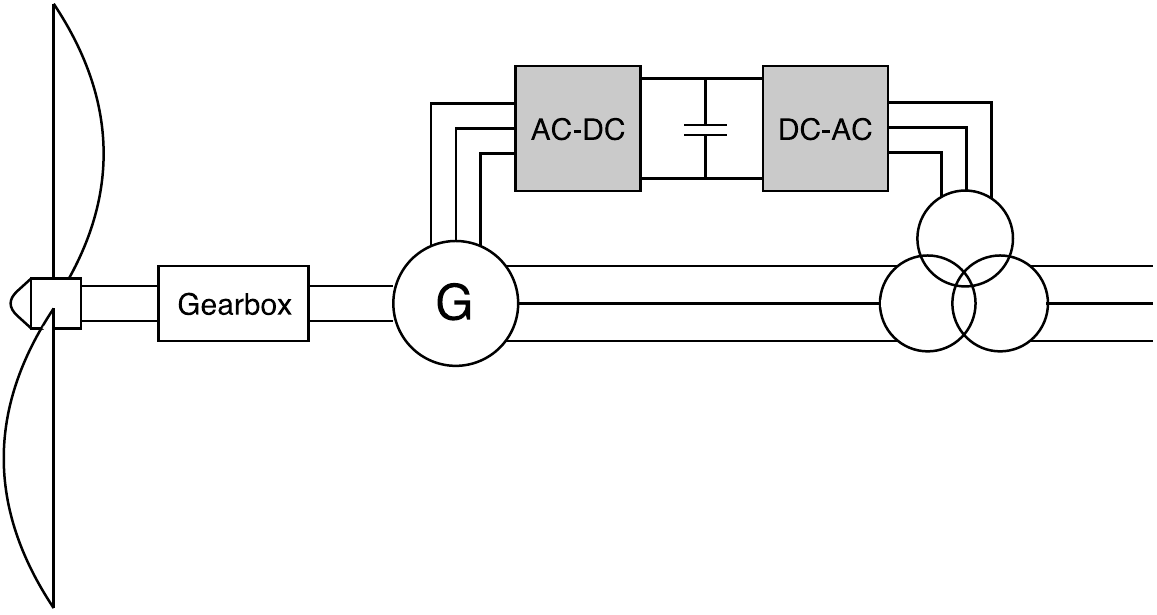}\label{fig.dfig}}
  \subfigure[Full-converter wind turbine. {\em{Type 4}}]{\includegraphics[width=0.35\textwidth]{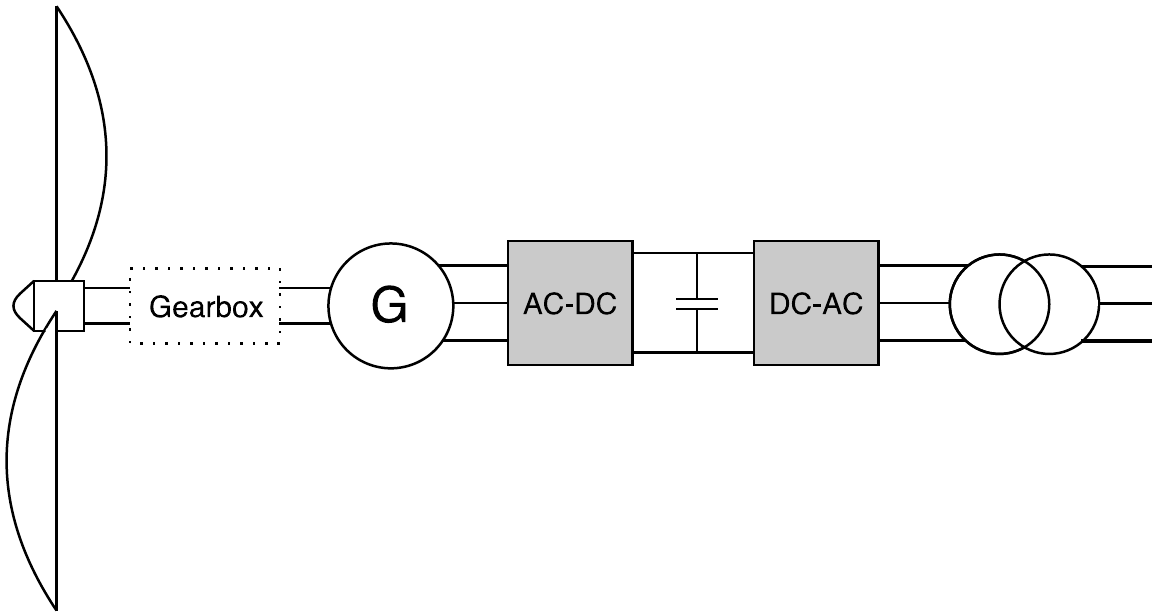}\label{fig.full}}
  \caption{Variable speed wind turbines.}\label{fig.vswt}
\end{figure}

WTs are usually classified as fixed speed wind turbines~(FSWTs) or
variable speed wind turbines~(VSWTs), see
Figure~{\ref{fig.clasification}}. FSWTs work at the same rotational
speed regardless of the wind speed. On the other hand, VSWTs can
operate around their optimum power point for each wind speed, using a
partial or full additional power converter, see Figure~{\ref{fig.vswt}}. As a result, VSWTs are
considered as more efficient solutions than FSWTs~\cite{njiri16}. With
the aim of providing generic electrical simulation models of wind
power generation, different standards have been developed such as the
IEC 61400--27--1.

\section{Simulation tool review}\label{sec.simulation_tool}

Various mathematical computational packages are available for research and educational purposes. Among the different solutions, Matlab is widely considered as the most often used software~\cite{sharma10}. Moreover, it is commonly recommended for electrical and electronics students, due to its user-friendly and easily understood by the students~\cite{vijayalakshmi19}. Matlab software combines a desktop environment tuned for iterative analysis and design processes with a programming language that expresses matrix and array mathematics directly, as stated in its website. However, as Matlab is commercial software, alternative open-source solutions have been developed in the last decades~\cite{kouroussis12}
The most popular free mathematical software alternatives to Matlab are the following: Freemat, Mathnium, Octave, R and Scilab. One of the most comprehensive study found in the literature review has been addressed by Glavelis {\em{et al.}}~\cite{glavelis10}, but excluding Matlab. 
Tables~\ref{tab.data_types}--\ref{tab.help} summarise a full comparison between Matlab and 
the FOSS alternatives. \textcolor{black}{These tables provide a complete characterisation by considering supported data types, language constructs, the functionality of the programming environment, help facilities and speed comparison analysis respectively.}
With regard to the computational time costs, Glavelis \emph{et al.}~\cite{glavelis10} conclude that the best average performance between the aforementioned FOSS is Scilab. Rubtsova and Korolev~\cite{rubtsova16} affirm that Scilab can be considered as the most complete open-source alternative to Matlab. Scilab also includes Xcos, a graphical editor of dynamic systems very similar to Simulink graphical editor in Matlab~\cite{arianto16}. Therefore, teaching material developed with Matlab/Simulink can be successfully replaced by equivalent material developed with Scilab/Xcos through some adjustments and moderate additional effort {\cite{1655368}}. {\color{black}{In conclusion, and considering the extended comparison depicted in Tables~\ref{tab.data_types}--\ref{tab.help} as
well as such contributions discussed in this Section, Scilab-Xcos is selected in this paper as the most suitable open-source software for our proposals.}} {\color{black}{Nevertheless, other studies can be found in the specific literature to help learners in Bachelor and Master Degree levels based on other free software solutions. For example, Qin {\em{et al}} {\cite{8974097}} recently show the initial efforts in the creation of OpenRES-an open-source JModelica.org library for renewable energy resources. Kanoj {\em{et al}} {\cite{6629933}} simulate an autonomous wind energy conversion system for irrigation purpose employing induction machines using Python.}}

\begin{table}[tbp]
	\small\sf\centering
	\caption{Supported data types}
	\resizebox{\linewidth}{!}{%
		\begin{tabular}{lllllll}
			\toprule
			Data type & Matlab & Freemat & Mathnium & Octave & R & Scilab\\
			\midrule
			Integer & X & X & X & X & X & X \\
			Double & X & X & X & X & X & X \\
			Boolean & X & X & X & X & X & X \\
			Complex & X & X & X & X & X & X \\
			Matrices & X & X & X & X & X & X \\
			Strings & X & X & X & X & X & X \\
			Structures & X & X & X & X & X & X \\
			Cells & X & X & X & X & X & X \\
			Java class & X &  & X &  &  &  \\
			Sparse matrix & X & X & X & X & X & X \\
			\bottomrule
		\end{tabular}
	}
	\label{tab.data_types}
\end{table}
\begin{table}[tbp]
	\small\sf\centering
	\caption{Supported language constructs}
	\resizebox{\linewidth}{!}{%
		\begin{tabular}{lllllll}
			\toprule
			Language constructs & Matlab & Freemat & Mathnium & Octave & R & Scilab\\
			\midrule
			Object-oriented & X & X & X &  & X & X \\
			Dynamic allocation of memory & X & X &  &  & X & X \\
			Sparse matrix & X & X & X & X & X & X \\
			Parallel programming & X & X &  & X & X & X \\
			\bottomrule
		\end{tabular}
	}
	\label{tab.language_constructs}
\end{table}
\begin{table}[tbp]
	\small\sf\centering
	\caption{Supported functionality of the programming environment}
	\resizebox{\linewidth}{!}{%
		\begin{tabular}{lllllll}
			\toprule
			Functionality & Matlab & Freemat & Mathnium & Octave & R & Scilab\\
			\midrule
			Debugger & X & X & X & X & X & X \\
			Profiler & X & X &  &  & X & X \\
			Syntax highlighting & X & X &  & X &  & X \\
			Creation of graphical user interface & X &  &  &  &  & X \\
			\bottomrule
		\end{tabular}
	}
	\label{tab.funcionality}
\end{table}
\begin{table}[tbp]
	\small\sf\centering
	\caption{Supported help facilities}
	\resizebox{\linewidth}{!}{%
		\begin{tabular}{lllllll}
			\toprule
			Help facilities & Matlab & Freemat & Mathnium & Octave & R & Scilab\\
			\midrule
			Help environment & X & X &  & X & X & X \\
			Manual & X & X &  & X & X & X \\
			Searching index & X & X &  &  & X & X \\
			Online documentation & X & X & X &  & X & X \\
			Mailing list/user forum & X & X &  & X & X & X \\
			\bottomrule
		\end{tabular}
	}
	\label{tab.help}
\end{table}

\section{Simulation set-up \& results}\label{sec.results}

As was previously discussed, the authors have proposed practical
sessions with Scilab-Xcos in the 'Wind energy' subject of
both the Electrical Engineering~BSc and Renewable Energy~MSc of
Universidad Polit\'{e}cnica de Cartagena (Spain). The VSWT model is based on the GE 3.6 model available in the
specific literature~\cite{miller03}. {\color{black}{The model corresponds to a \emph{type 3} WT, though the proposed simulation can be also applied to a \emph{type 4} WT, as stated in~\cite{ullah08}.}} 
Source codes and data models are provided to the students. Actually, files are also available to other researchers and students through the following link {\url{https://n9.cl/3hr0z}}.
{\color{black} Additional analysis can be also proposed to extend these simulations toward a more complex study, mainly focused on post-graduate students. For example, a more complex wind turbine model discussed by the authors in {\cite{en12091631}} can be used to analyse grid code compliance. In a similar way, more realistic wind speed profiles characterised in {\cite{8656477}} can be also considered for simulations.}

\subsection{Scilab simulation}\label{sec.scilab}

Based on the different \emph{scripts} prepared by the authors,
students can graphically analyse the relationship among the main
parameters of wind turbines (previously introduced in theoretical
sessions): $\beta$, $\lambda$, $v_{w}$, $C_{p}$, $P_{mech}$ and
$\Omega_{WT}$. By using the Scilab-Xcos \emph{scripts}, the
students can compare their graphical results facing the estimated VSWT
curves. These curves are following described. Results are also
included as examples of our practical sessions.
\begin{enumerate}
\item[(i)] $C_{p}-v_{w}$ curve: 
  The rotor performance of a WT is characterised by its power
  coefficient $C_{p}$, having a limit of $\frac{16}{27}\approx 59\%$
  according to the Betz limit. The power
  coefficient $C_{p}$ can be estimated depending on the pitch angle
  $\beta$ and the tip speed ratio $\lambda$: 
  \begin{equation}
    \label{eq.cp}
    C_{p}(\lambda,\beta)=\sum_{i=0}^{4}\sum_{j=0}^{4}\alpha_{i,j}\beta^{i}\lambda^{j} ,
  \end{equation}
  where coefficients $\alpha_{i,j}$ are taken from Table~\ref{tab.coeficientes}. The tip speed ratio~$\lambda$ is defined as 
	\begin{equation}    
    \label{eq.lambda}   
    \lambda=\dfrac{\Omega_{0}\cdot R\cdot\Omega_{WT}}{V_{W}} ,
    \end{equation}
\textcolor{black}{where $\Omega_{0}$ is the rotor base speed (rad/s), $\Omega_{WT}$
refers to the rotor speed (pu), $R$ is the rotor radius (m) and
$v_{w}$ is the wind speed (m/s).} According to Miller {\em{et
          al.}}~\cite{miller03}, the rotor reference speed is normally
      1.2~pu, assuming that $\Omega_{WT}=1.2$. The pitch angle can
      vary between 0 and 27$^{\circ}$. Cut-in and cut-out wind speeds
      for this WT model are 4~$m/s$ and 25~$m/s$, respectively. 
      With all these considerations, the $C_{p}-v_{w}$ curve depending on
      the pitch angle can be estimated as depicted in Figure~\ref{fig.Cp-Vw}.
	\begin{figure}[tbp]
		\centering
		\includegraphics[width=\linewidth]{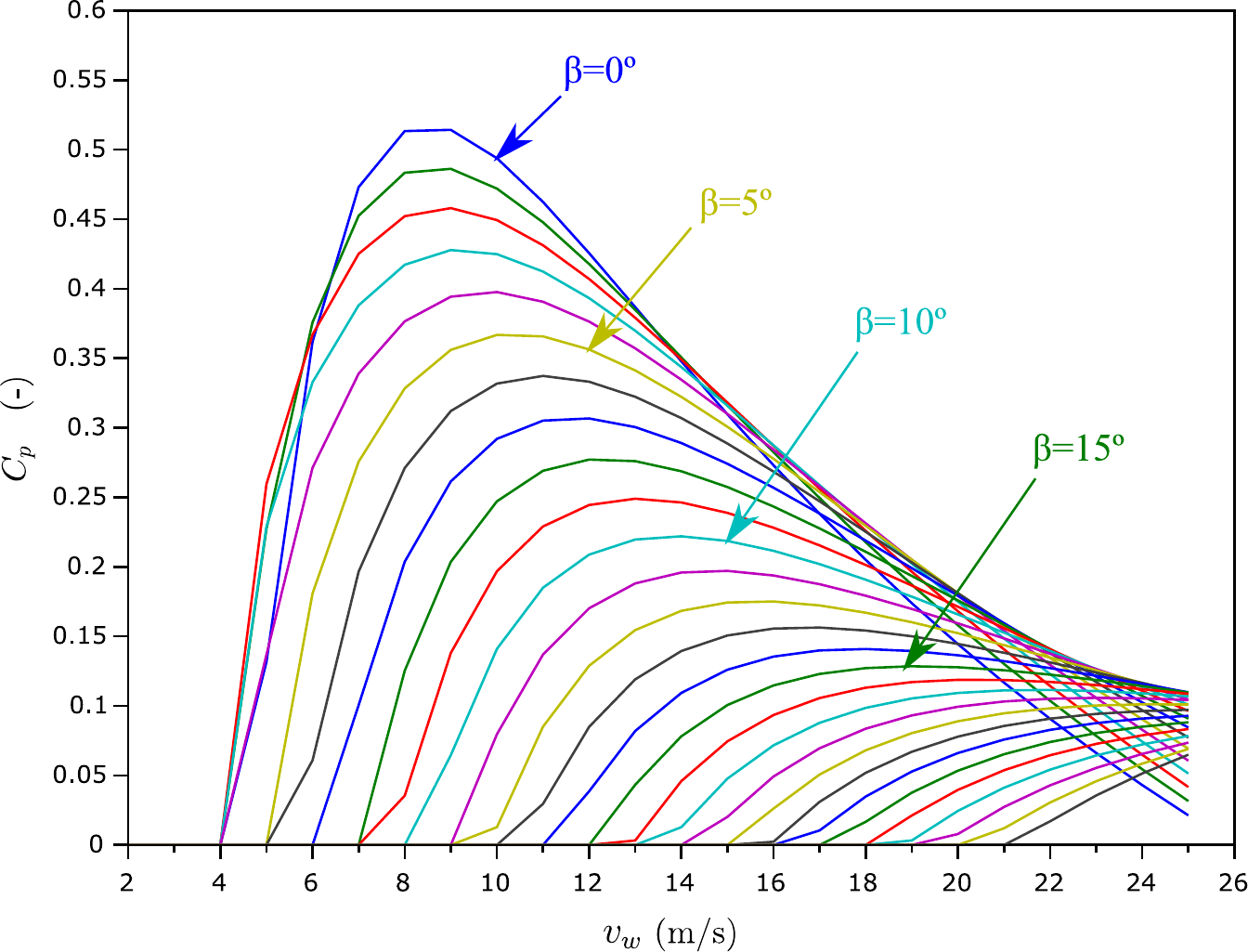}
		\caption{$C_{p}=f(v_{w},\beta)$. Scilab: \texttt{plot(Vw,Cp)} }
		\label{fig.Cp-Vw}
	\end{figure}
	From these results, 
        students can graphically check that the higher pitch angle
        $\beta$, the lower power coefficient. These lower power coefficient values also reduce the shaft mechanical power, as explained afterwards.
      \item[(ii)] $C_{p}-\lambda$ curve: According to the specific
        literature, $C_{p}$ is usually considered as a function of the
        tip speed ratio~$\lambda$
        Moreover, the $C_{p}-\lambda$ curve is necessary for developing the turbine control and guaranteeing a safe operation of the WT. 
        According to the definition of $\lambda$ in eq.~\eqref{eq.lambda}, and considering a constant wind speed value, if the WT rotational speed is low, too much wind will pass undisturbed through the blades. On the contrary, if the rotational speed value is high ($\lambda\approx13$), the rotating blades appear as a 'solid disc' to the wind, creating a large amount of drag~\cite{ragheb11}. 
        %
        %
        %
        %
        In practice, the specific
        literature suggests that the optimum $\lambda$ for a
        three-bladed WT is generally estimated as $\lambda_{opt}\approx7$~\cite{ata10}. In the model under
        consideration, it is suggested that
        $2\leq\lambda\leq13$. 
        Figure~\ref{fig.Cp-lambda} shows the $C_{p}-\lambda$ 
        curve depending on the pitch angle. 
	\begin{figure}[tbp]
          \centering
          \includegraphics[width=\linewidth]{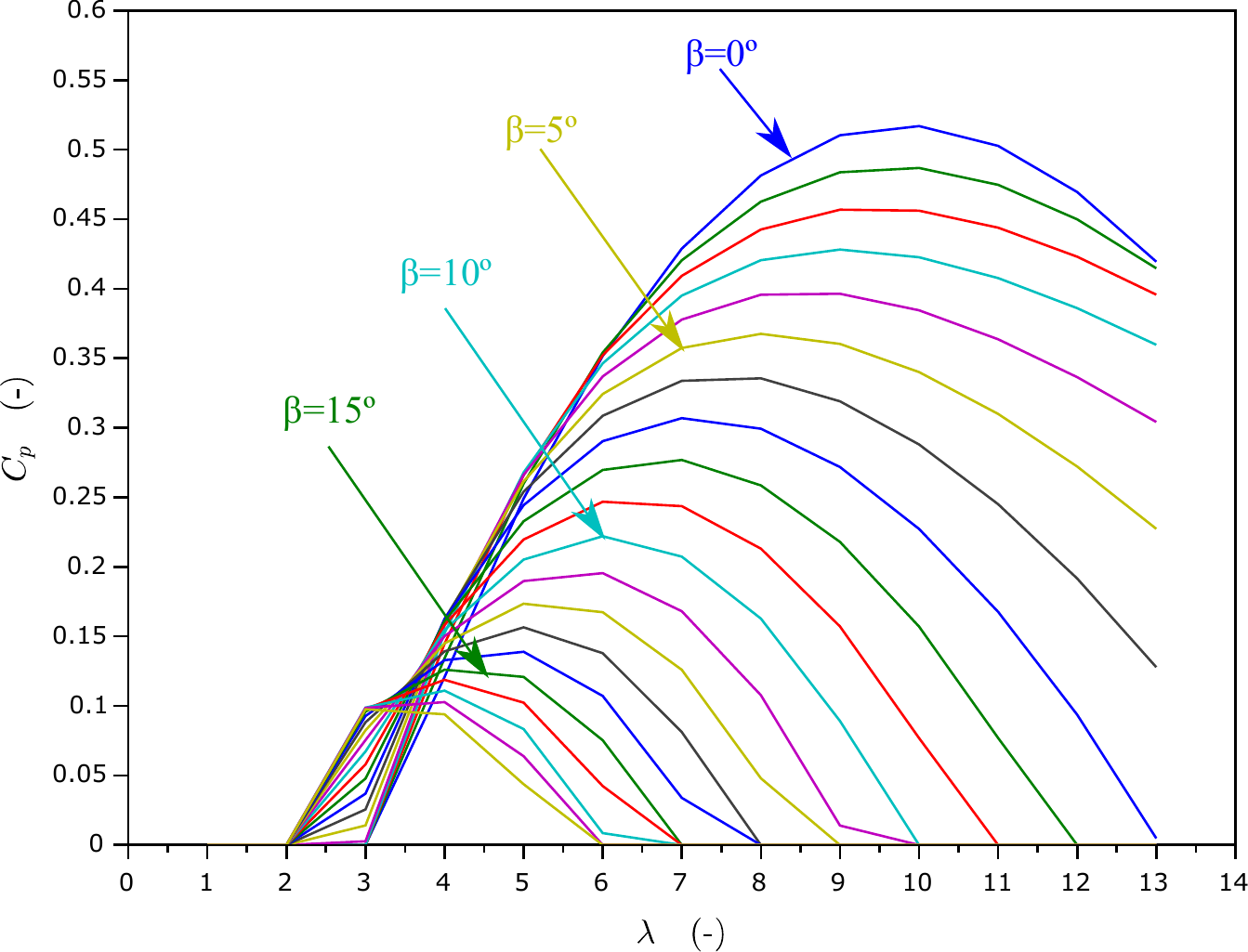}
          \caption{$C_{p}=f(\lambda,\beta)$. Scilab: \texttt{plot(lambda,Cp)}}
          \label{fig.Cp-lambda}
        \end{figure}
        According to the results and the theoretical sessions given to the students, the larger pitch angle $\beta$ value, 
        the lower power coefficient. Moreover, it can be determined an optimum value of the tip speed
        ratio~$\lambda_{opt}$ for each $\beta$ value ---which addresses an optimum $C_{p}$ value accordingly---. 
        Due to the VSWT power converter performances, they are able to
        work on each optimum $C_{p}^{{\tiny{Opt}}}$ value for each wind
        speed. \textcolor{black}{This point is a spotlight that should be clarified and
        highlighted for the students.}        
      \item[(iii)] $P_{mech}-\Omega_{WT}$ curve: 	The mechanical power $P_{mech}$ is obtained from:
	\begin{equation}
\label{eq.pmt}
P_{mech}=K_{rotor}\cdot C_{P} \cdot v_{w}^{3}\,,
\end{equation}
being $K_{rotor}=\frac{1}{2}\cdot\frac{1}{S_{n}}\cdot \rho\cdot A_{r}$ ($S_{n}$ the rated power, $\rho$ the air density, $A_{r}$ the swept area by the blades), $C_{P}$ the power coefficient, and $v_{w}$ the wind speed. The power coefficient $C_{P}$ is estimated with eq. \eqref{eq.cp}. By considering $K_{rotor}=0.00145$, $P_{mech}$ is obtained in pu on the MW base of the wind turbine.  
In line with the literature review, $P_{mech}$ is commonly plotted vs the rotor rotational speed  $\Omega_{WT}$, 
which can vary between 0 and the corresponding reference value (1.2 pu). Two different options are then proposed to analyse the $P_{mech}-\Omega_{WT}$ curve:
  
  \begin{table}[tbp]
    \small\sf\centering
		\caption{Coefficients $\alpha_{i,j}$ to calculate
			$C_{p}(\lambda,\beta)$}
		\resizebox{\linewidth}{!}{%
			\begin{tabular}{llllll}
				\toprule
				i,j & 0&1&2&3&4\\
				\midrule
				0 &$-4.19\cdot10^{-1}$&$2.18\cdot10^{-1}$&$-1.24\cdot10^{-2}$&$-1.34\cdot10^{-4}$&$1.15\cdot10^{-5}$\\
				1 &$-6.76\cdot10^{-2}$&$6.04\cdot10^{-2}$&$-1.39\cdot10^{-2}$&$1.07\cdot10^{-3}$&$-2.39\cdot10^{-5}$\\
				2 &$1.57\cdot10^{-2}$&$-1.01\cdot10^{-2}$&$2.15\cdot10^{-3}$&$-1.49\cdot10^{-4}$&$2.79\cdot10^{-6}$\\
				3& $-8.60\cdot10^{-4}$&$5.71\cdot10^{-4}$&$-1.05\cdot10^{-4}$&$5.99\cdot10^{-6}$&$-8.91\cdot10^{-8}$\\
				4 &$1.48\cdot10^{-5}$&$-9.48\cdot10^{-6}$&$1.62\cdot10^{-6}$&$-7.15\cdot10^{-8}$&$4.97\cdot10^{-10}$\\
				\bottomrule
			\end{tabular}
		}
		\label{tab.coeficientes}
	\end{table}

	\begin{enumerate}
        \item {\em{Constant pitch angle ($\beta$) and variable wind
              speed ($v_{w}$)}}. $\beta$ is fixed at 0$^\circ$ and the
          wind speed is gradually increased from 4 to 12~$m/s$ for the sake of
          clarity of the figure, see~Figure~\ref{fig.Pmec_omega_1}.
          $P_{mech}-\Omega_{WT}$ curves then reach their maximum
          values at the optimum power coefficient according to results
          depicted in Figure~\ref{fig.Cp-lambda}. Moreover, due to the
          cubic dependence between $v_{w}$ and $P_{mech}$, such
          mechanical power run up as wind speed increases (please
          compare, for instance, $P_{mech}$ at $v_{w}=4$~$m/s$ and
          $v_{w}=8$~$m/s$).
		\begin{figure}[tbp]
			\centering
			\includegraphics[width=\linewidth]{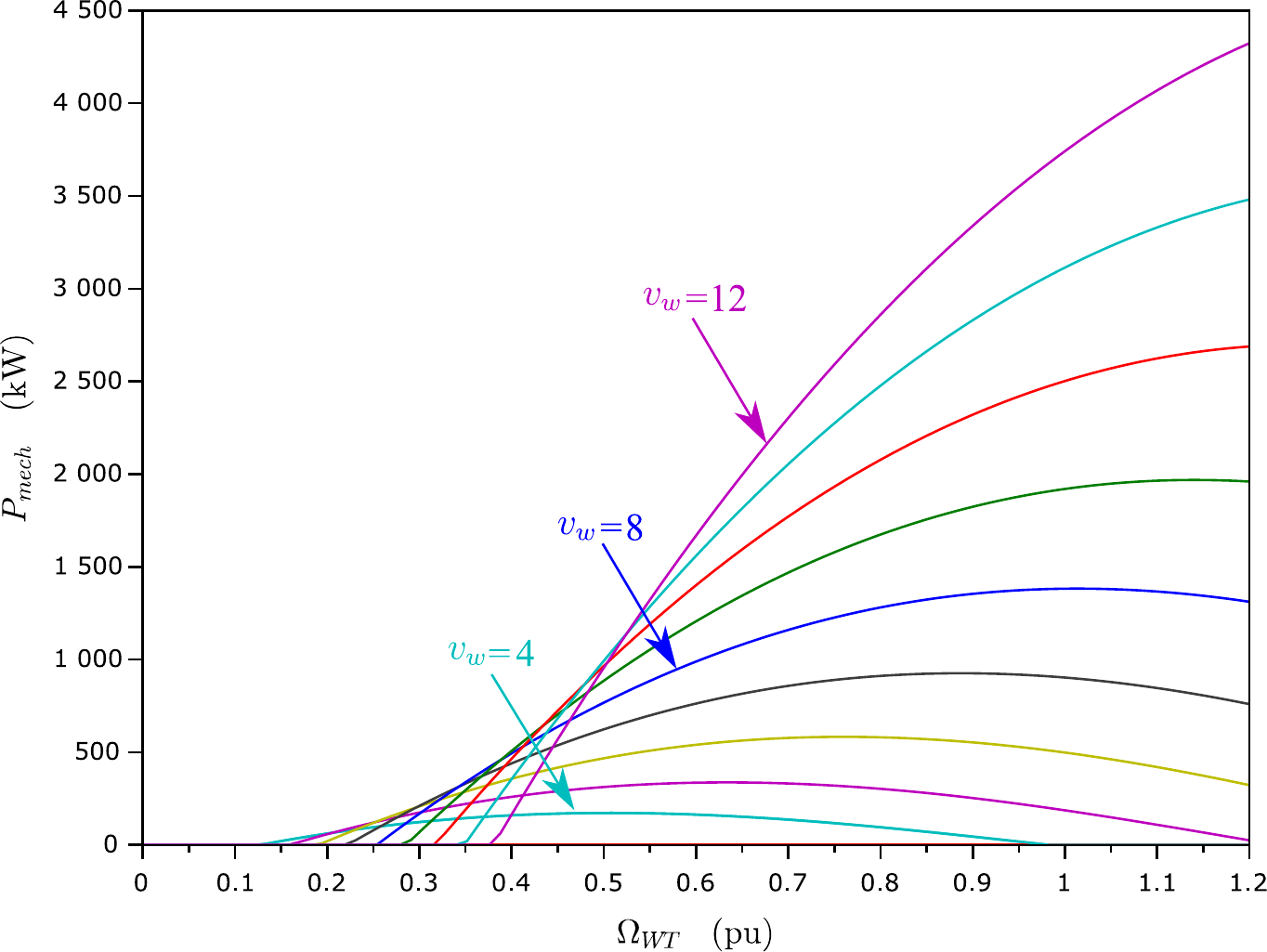}
			\caption{$P_{mech}=f(\Omega_{WT},v_{w})$. Scilab: \texttt{plot(w,Pmec)}}
			\label{fig.Pmec_omega_1}
		\end{figure}		
              \item {\em{Constant wind speed ($v_{w}$) and variable
                    pitch angle ($\beta$)}}. In this case, wind speed
                is fixed at $v_{w}=12$~$m/s$ and $\beta$ is increased
                from 0 to 19$^\circ$. Figure~\ref{fig.Pmec_omega_2}
                shows the results obtained under this assumption. As in
                previous cases, a $\beta$ increasing reduces the
                mechanical power due to the power coefficient
                decreasing. In our opinion, it is interesting for the
                students to estimate the mechanical power differences
                between $\beta=0^\circ$ and $\beta=19^\circ$ ---which
                correspond $\approx3250$~kW for the selected
                WT---. These results allow the student to evaluate in
                detail the severe mechanical power $P_{mech}$
                dependence on the pitch angle.
		\begin{figure}[tbp]
			\centering
			\includegraphics[width=\linewidth]{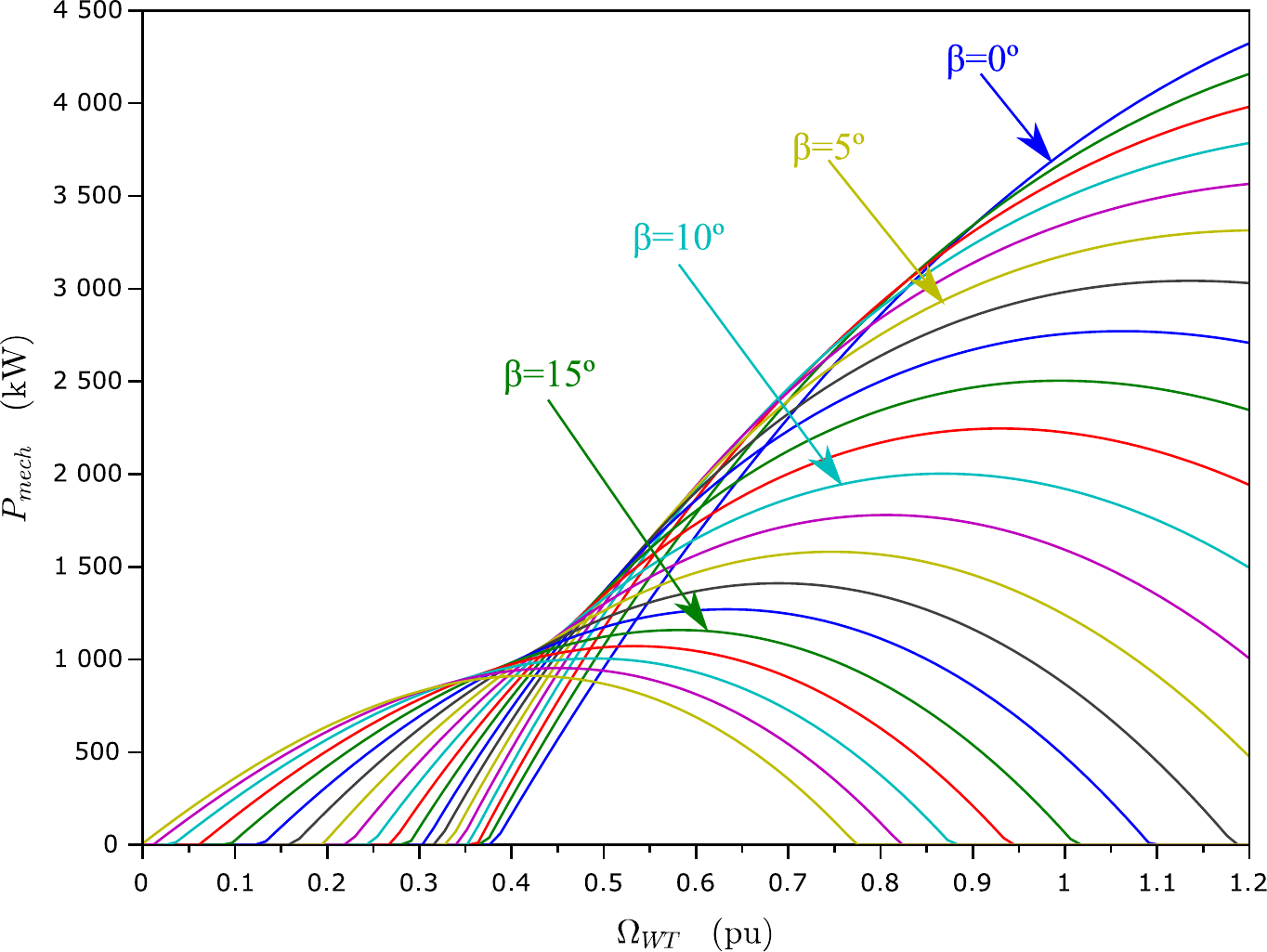}
			\caption{$P_{mech}=f(\Omega_{WT},\beta)$. Scilab: \texttt{plot(w,Pmec)}}
			\label{fig.Pmec_omega_2}
		\end{figure}		
	\end{enumerate}
\end{enumerate}

\subsection{Xcos simulation}\label{sec.xcos}

\begin{figure}[tbp]
	\centering
	\includegraphics[width=\linewidth]{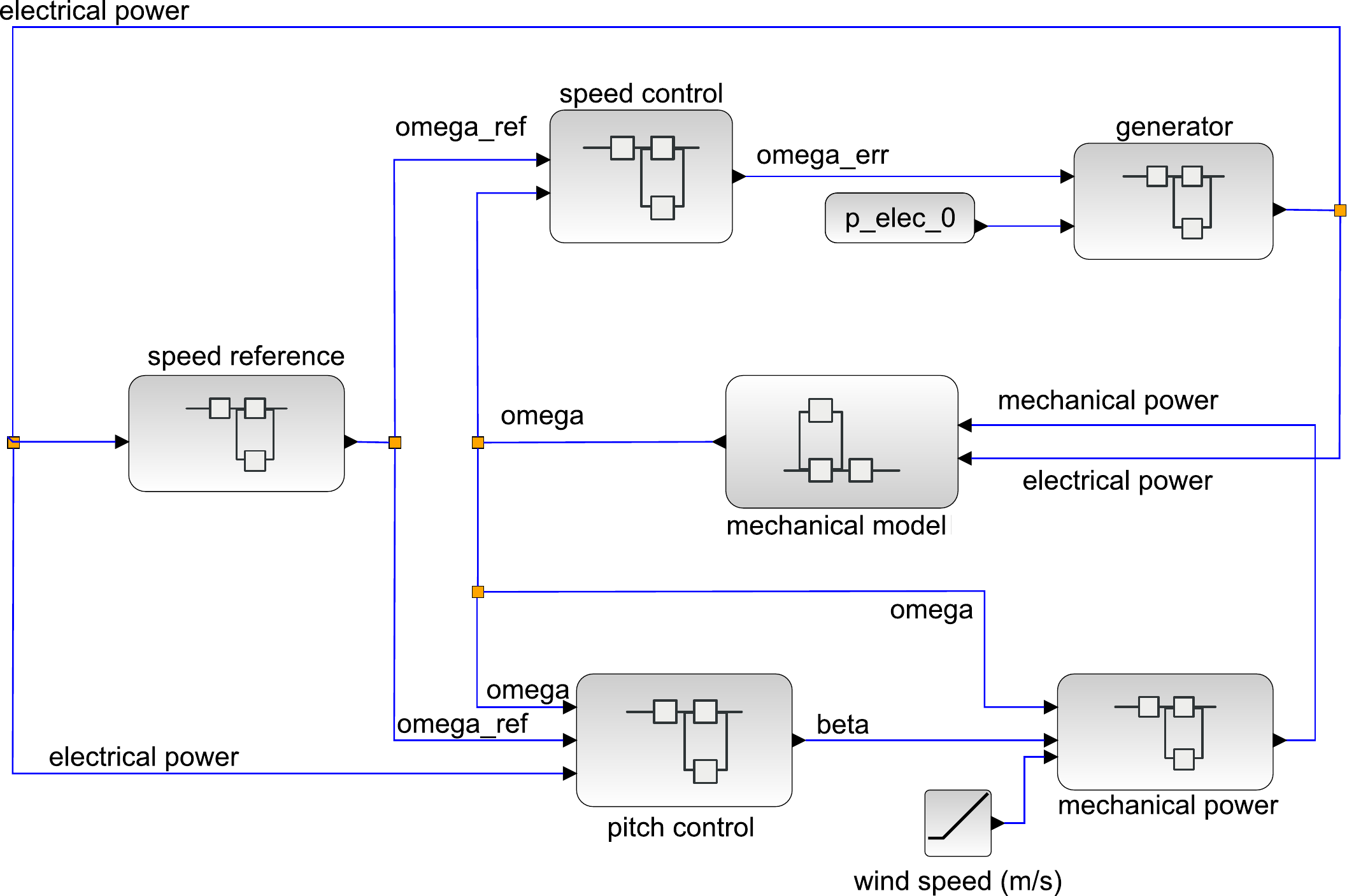}
	\caption{\textcolor{black}{Wind turbine block model. Scilab--Xcos environment.}}
	\label{fig.aero}
\end{figure}

\begin{figure}[tbp]
	\centering
	\includegraphics[width=\linewidth]{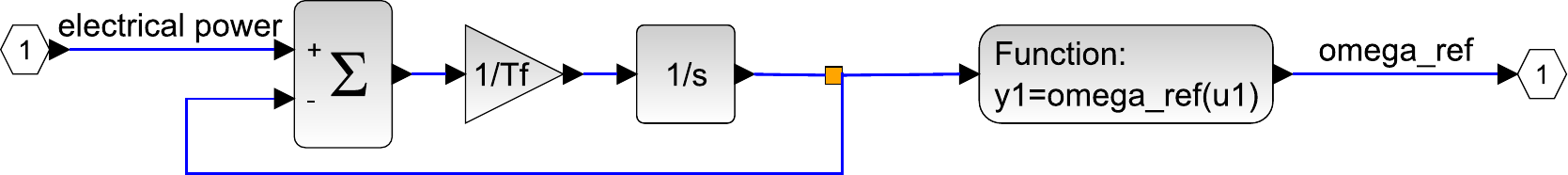}
	\caption{\textcolor{black}{Rotational speed reference block diagram. Scilab--Xcos environment.}}
	\label{fig.omega_ref}
\end{figure}

As was previously discussed, the selected VSWT has been simulated by
the authors by using the Scilab-Xcos open-source
solution. Apart from the parameters previously analysed ($\beta$,
$\lambda$, $v_{w}$, $C_{p}$, $P_{mech}$ and $\Omega_{WT}$), the
graphical {\texttt{Xcos}} environment allows us to determine the
electrical power $P_{e}$ provided by the WT and the generator
rotational speed $\Omega_{g}$. The block diagram of the VSWT simulated
within this graphical environment is depicted in
Figure~\ref{fig.aero}.  A comparison between two different mechanical
models of the rotor is also carried out in our practical sessions:
$(i)$~One-mass mechanical model and $(ii)$ Two-mass mechanical
model. {\color{black}{Therefore, the block labelled as 'Mechanical model' in
Figure~\ref{fig.aero} corresponds to both approaches. This block is then modified depending on the mechanical model under consideration.}} 
\textcolor{black}{The reference rotational speed $\Omega_{ref}$ is estimated from the measured active power $P_{ef}$:
\begin{equation}\label{eq.omegaref1}
\Omega_{ref}=-0.67\cdot P_{ef}^{2}+1.42\cdot P_{ef}+0.51 ,
\end{equation}
being $P_{ef}$ the measured electrical power $P_{e}$ after a delay $T_{f}$. The block diagram is depicted in Figure \ref{fig.omega_ref}. When the electrical measured power $P_{ef}$ is below $0.75$~pu, the rotational speed reference is calculated according to eq.~\eqref{eq.omegaref1}. If $P_{ef}$ is over $0.75$~pu, the pitch controller must keep constant the rotational speed at $\Omega_{ref}=1.2$~pu. }

\begin{figure}[tbp]
	\centering
	\includegraphics[width=\linewidth]{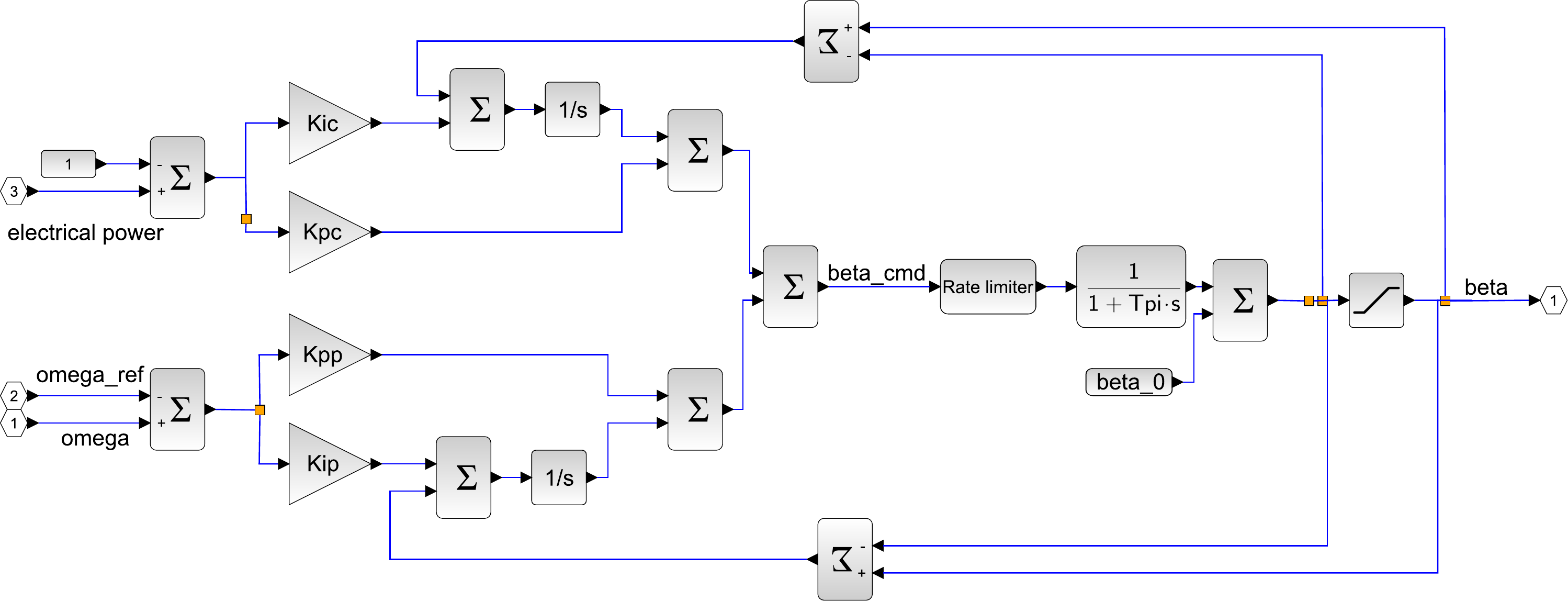}
	\caption{\textcolor{black}{Pitch angle control block diagram. Scilab--Xcos environment.}}
	\label{fig.pitch}
\end{figure}

\begin{figure}[tbp]
	\centering
	\includegraphics[width=\linewidth]{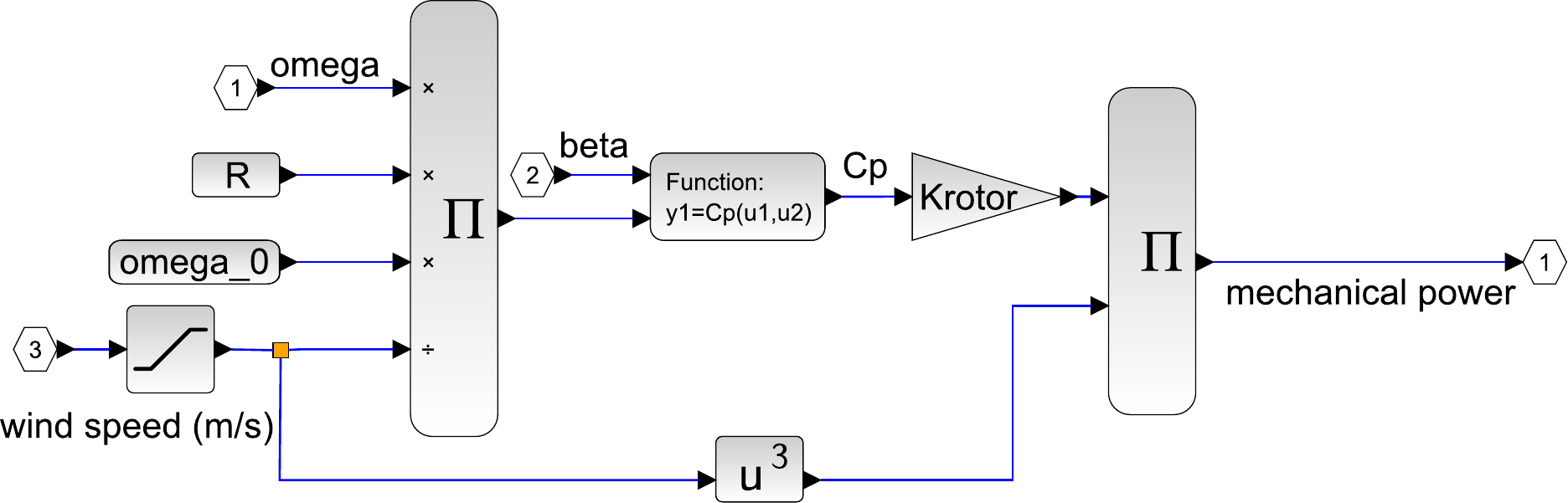}
	\caption{\textcolor{black}{Mechanical power block diagram. Scilab--Xcos environment.}}
	\label{fig.p_mec}
\end{figure}

\begin{figure}[tbp]
	\centering
	\includegraphics[width=\linewidth]{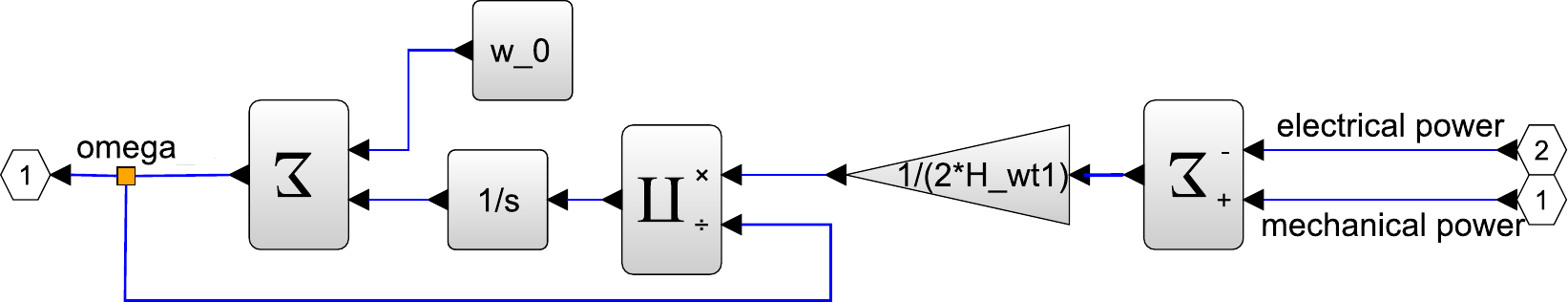}
	\caption{\textcolor{black}{One-mass mechanical model block diagram. Scilab--Xcos environment.}}
	\label{fig.one_mass}
\end{figure}

\textcolor{black}{The blades are pitched to limit the maximum mechanical power delivered
to the shaft. It is carried out by a combination of two
controllers~\cite{martinez18}: 
$(i)$ traditional pitch control and
$(ii)$ pitch compensation, see Figure~\ref{fig.pitch}. Both
controllers are modelled through conventional PI controllers, and including an anti windup. In reference to the pitch angle controller, the input variable to such PI controller is the difference between the actual rotor rotational speed and the reference rotor speed ($\Omega_{err}=\Omega_{WT} - \Omega_{ref}$). Regarding the pitch compensation PI controller, the input is determined from the difference between the electrical power $P_{e}$ and the maximum power $P_{max}=1$.} \textcolor{black}{The power coefficient and, subsequently, the mechanical power $P_{mech}$ can be estimated by using eq.~\eqref{eq.cp} and \eqref{eq.pmt}, respectively. The block diagram to calculate $P_{mech}$ with Xcos is depicted in Figure \ref{fig.p_mec}.} The WT rotational speed $\Omega_{WT}$ is estimated depending on the mechanical model. In this way, if the equivalent one-mass model is considered for simulations, the inertia equation provides $\Omega_{WT}=\Omega_{g}$, 
\begin{equation}\label{eq.momentos}
  \Omega_{WT}(s)=\dfrac{P_{e}(s)-P_{mech}(s)}{2 H_{WT}\cdot s},
\end{equation}
being $H_{WT}$ the WT equivalent inertia constant. \textcolor{black}{Figure \ref{fig.one_mass} shows the one-mass mechanical model block diagram implemented in Scilab--Xcos.}

\begin{figure}[tbp]
  \centering
  \subfigure[Two-mass model scheme]{\includegraphics[width=.7\linewidth]{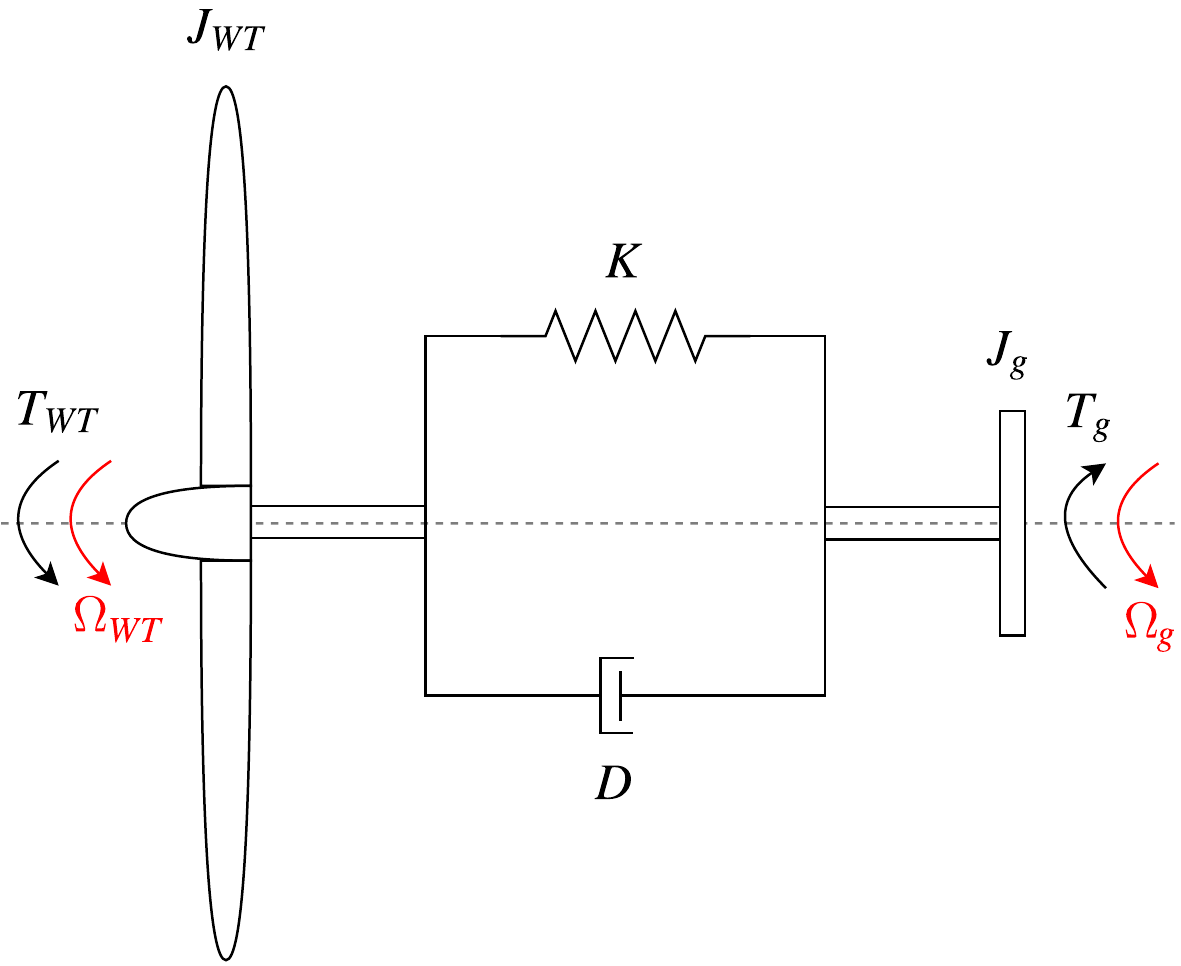}}
  \subfigure[\textcolor{black}{Two-mass block diagram model}]{\includegraphics[width=\linewidth]{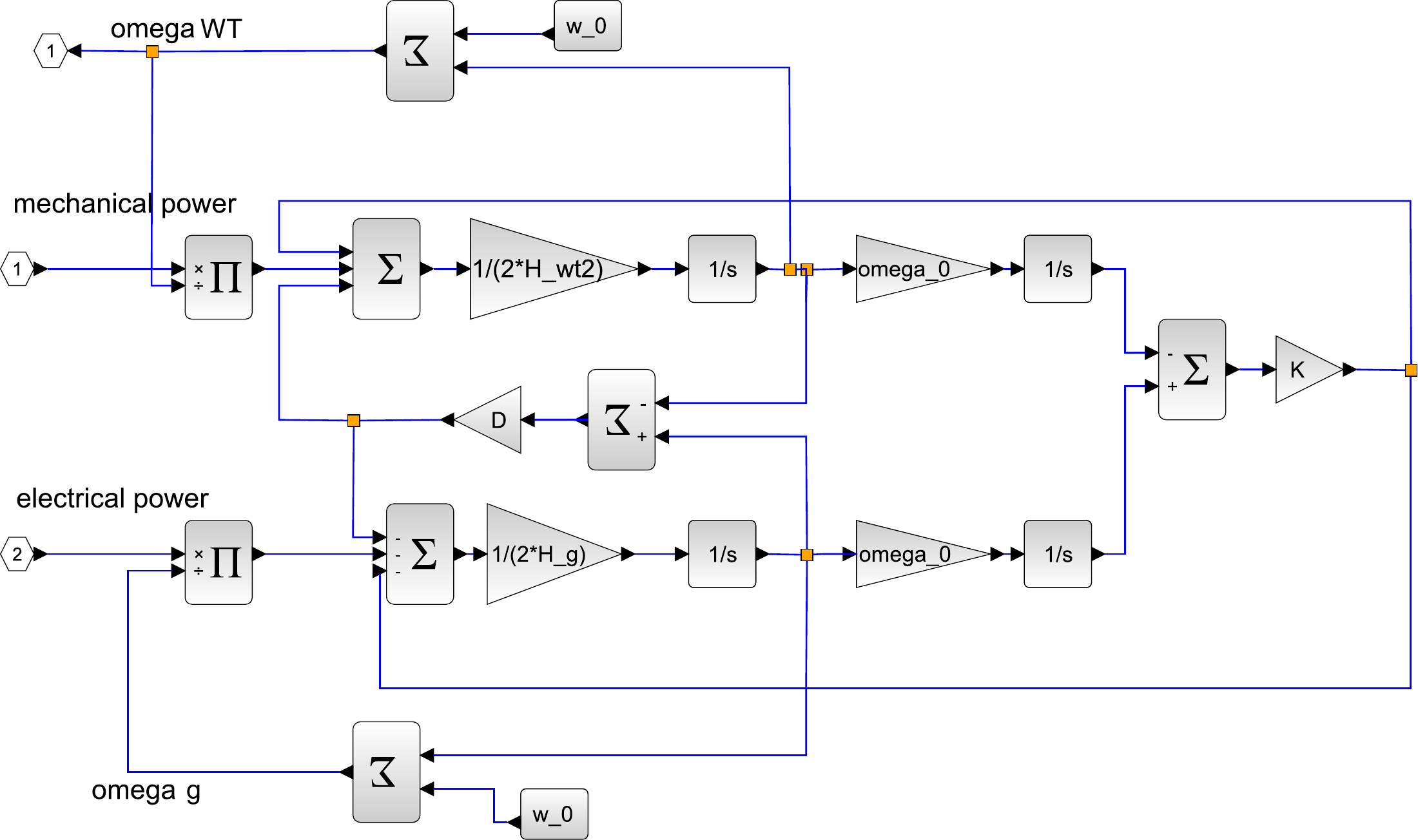}}
  \caption{Two mass model: scheme and block diagram. Scilab--Xcos environment.}
  \label{fig.two_mass}
\end{figure}

With regard to the two-mass model approach, it assumes both rotor and blades as
a single mass, and the generator as another mass, 
see Figure~\ref{fig.two_mass}. Hence, two different rotational speeds
are obtained: $(i)$~rotor rotational speed $\Omega_{WT}$ and
$(ii)$~generator rotational speed $\Omega_{g}$.  

\begin{figure}[tbp]
	\centering
	\includegraphics[width=\linewidth]{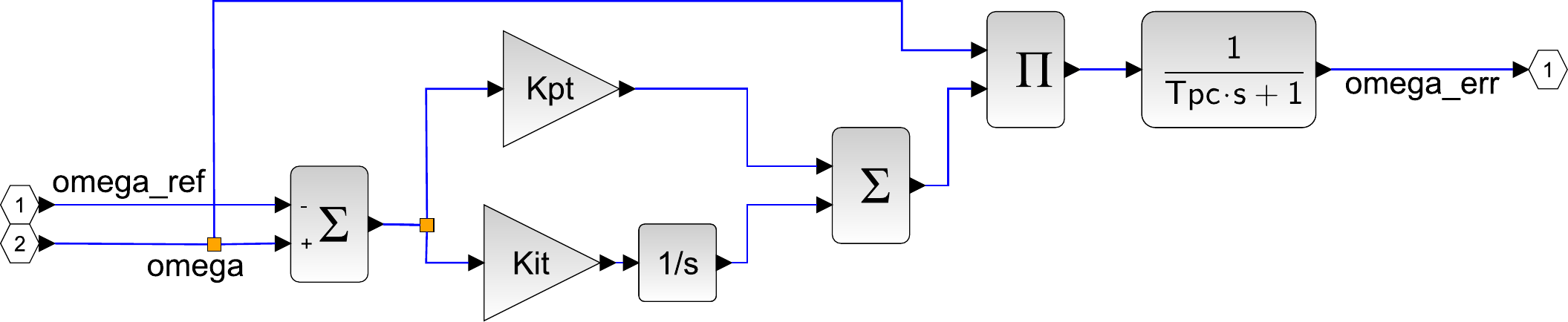}
	\caption{\textcolor{black}{Speed control block diagram. Scilab--Xcos environment.}}
	\label{fig.speed_control}
\end{figure}

\begin{figure}[tbp]
	\centering
	\includegraphics[width=\linewidth]{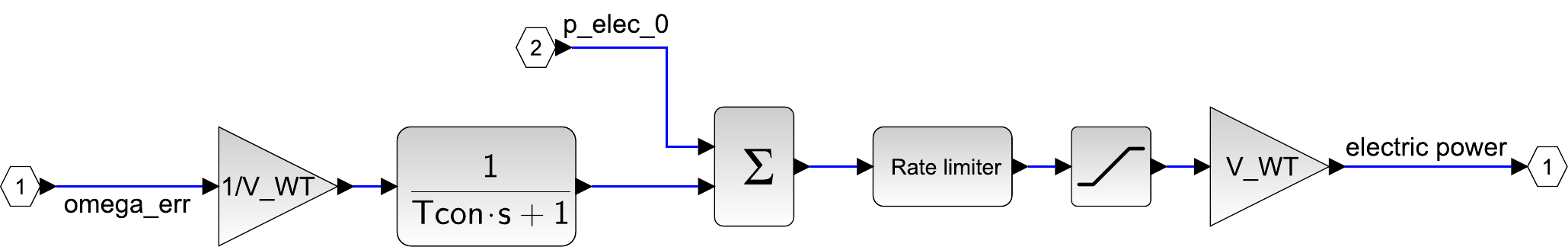}
	\caption{\textcolor{black}{Generator block diagram. Scilab--Xcos environment.}}
	\label{fig.generator}
\end{figure}

\textcolor{black}{The speed control block estimates the difference between the current rotational speed ($\Omega$) and the reference rotational speed value ($\Omega_{ref}$) through a PI controller, as depicted in Figure \ref{fig.speed_control}. This error ($\Omega_{err}$) is sent to the generator and power converter, determining the electrical power variation due to such rotational speed error ($\Omega_{err}$). A subsequent additional electric power value is then incorporated to the initial electric power value ($p\_elec\_0$), as depicted in Figure~\ref{fig.generator}.}

\begin{figure}[tbp]
  \centering
  \subfigure[Wind speed variation]{\includegraphics[width=0.4\textwidth]{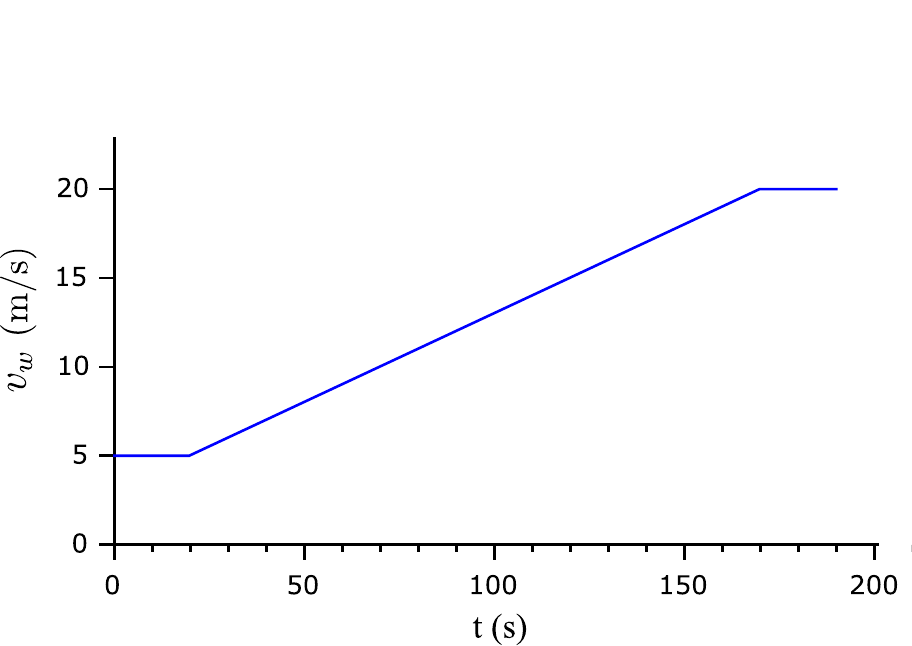}\label{fig.wind1}}
  \subfigure[Pitch angle]{\includegraphics[width=0.4\textwidth]{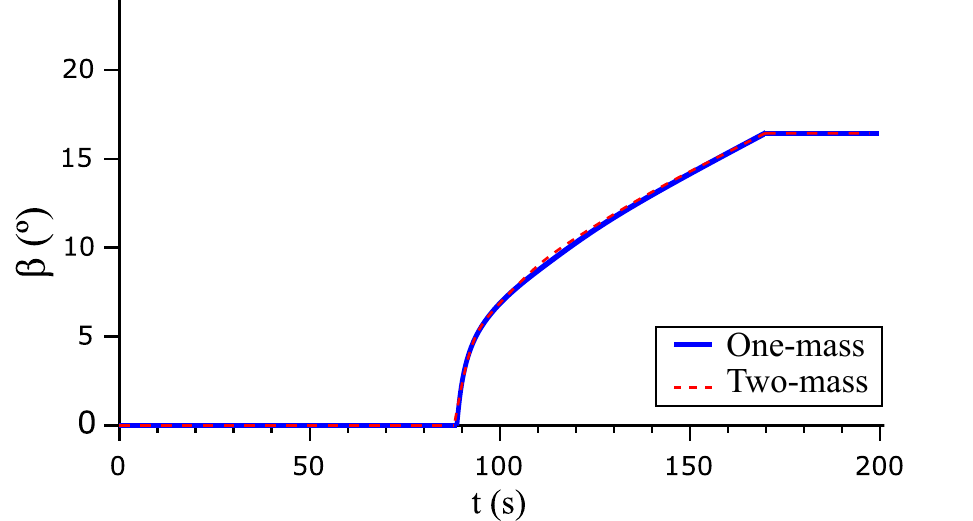}\label{fig.pitch1}}
  \subfigure[Tip speed ratio]{\includegraphics[width=0.4\textwidth]{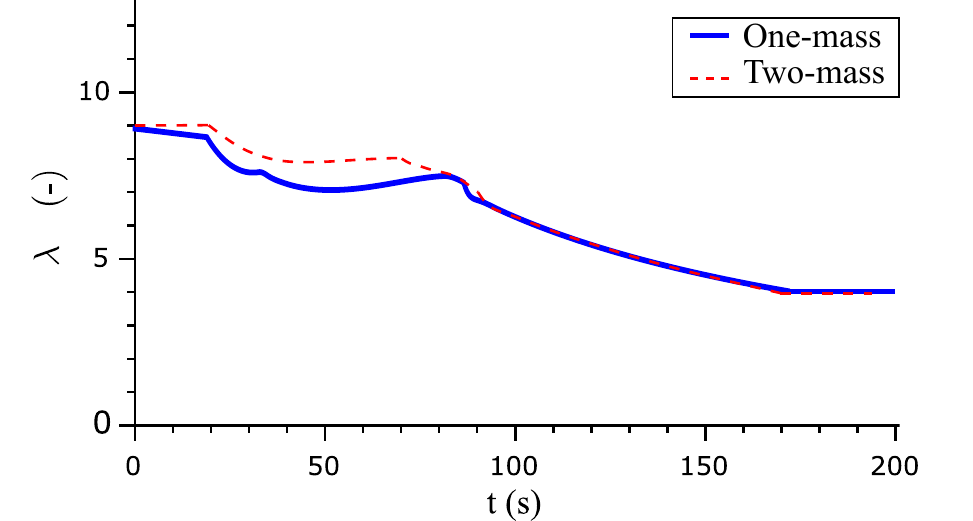}\label{fig.tsr1}}
  \subfigure[Power coefficient]{\includegraphics[width=0.4\textwidth]{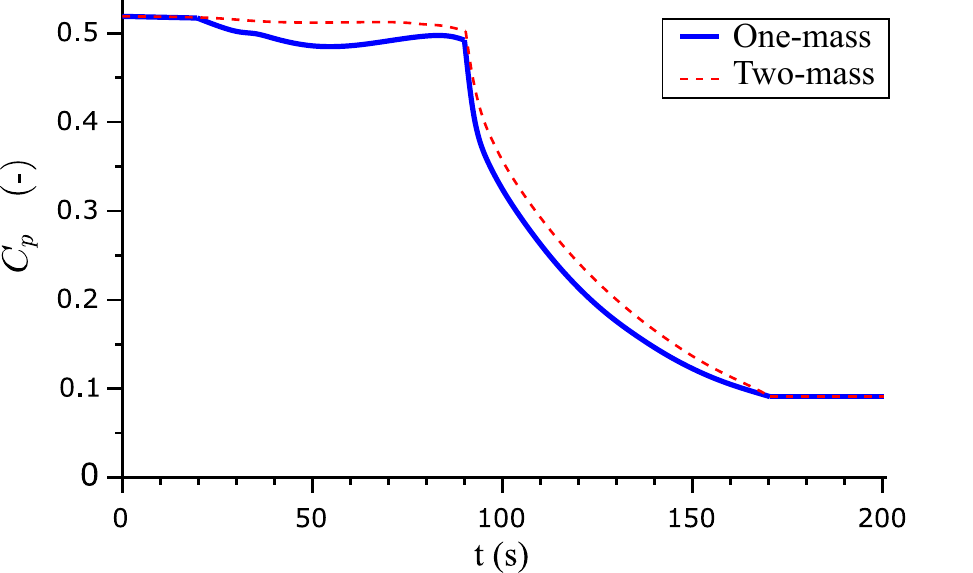}\label{fig.cp1}}
  \caption{Wind speed variation, pitch angle, tip speed ratio and power coefficient results for one-mass and two-mass mechanical models}
  \label{fig.results1}
\end{figure}

\begin{figure}[tbp]
  \centering
  \subfigure[Turbine rotational speed]{\includegraphics[width=0.4\textwidth]{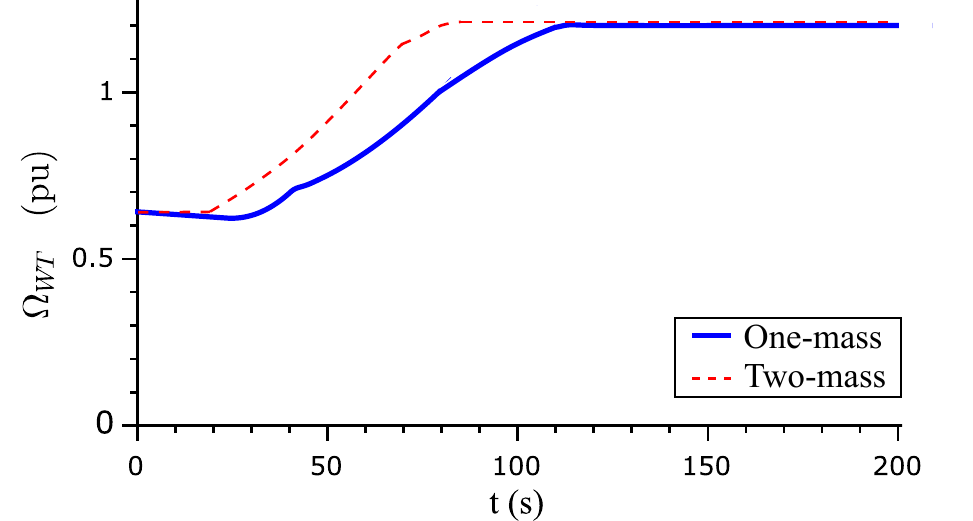}\label{fig.omega1}}
  \subfigure[Generator rotational speed]{\includegraphics[width=0.4\textwidth]{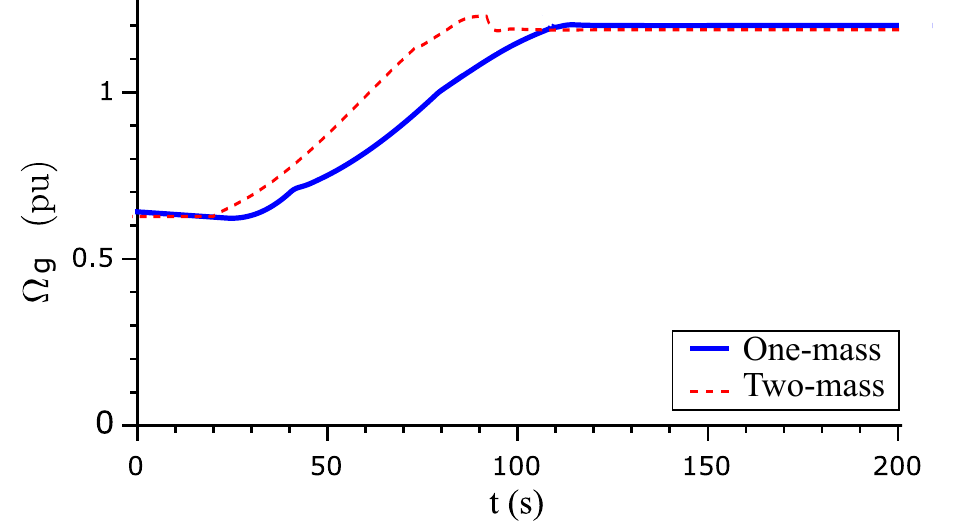}\label{fig.omega_G1}}
  \subfigure[Mechanical power]{\includegraphics[width=0.4\textwidth]{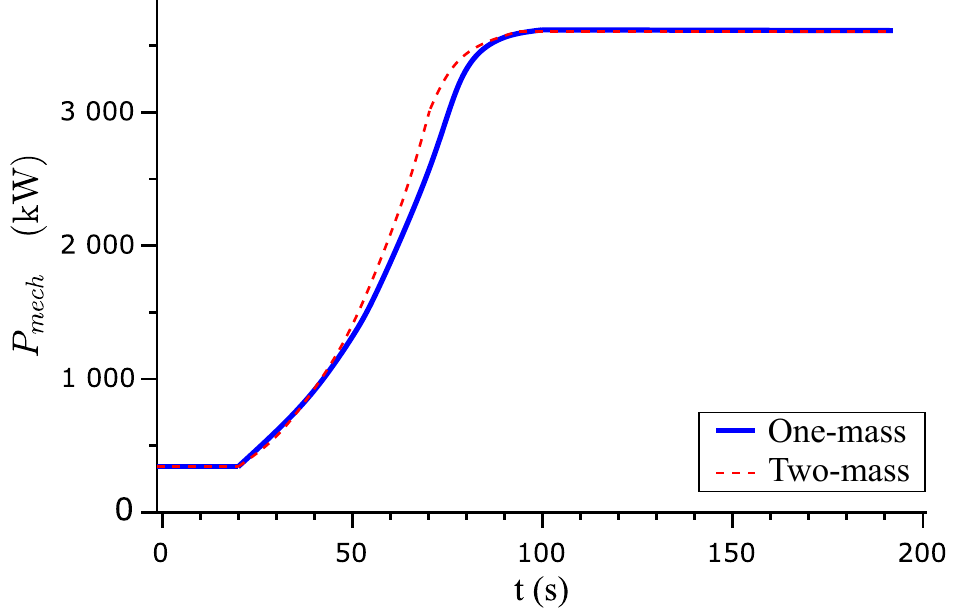}\label{fig.pm1}}
  \subfigure[Electrical power]{\includegraphics[width=0.4\textwidth]{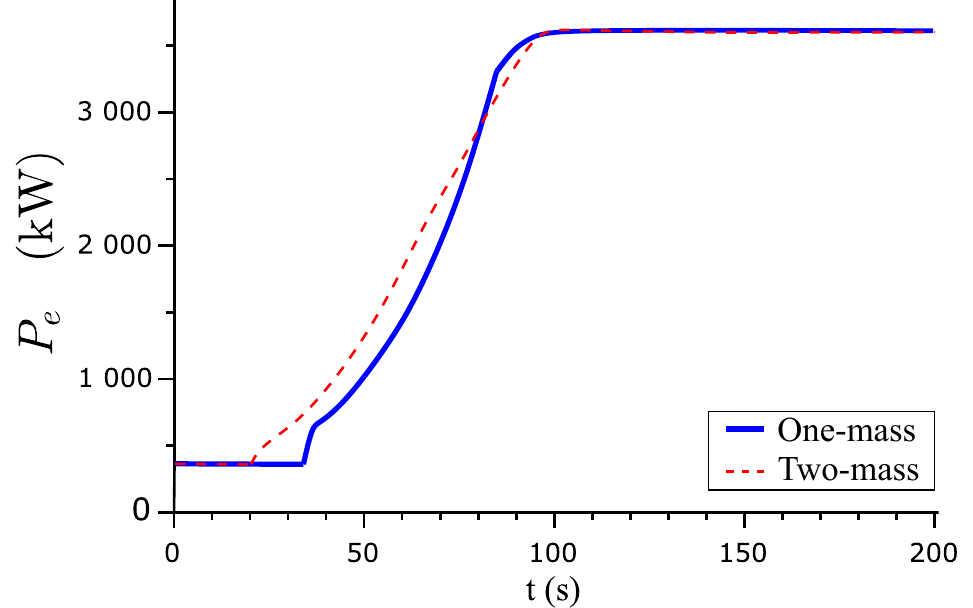}\label{fig.pe1}}
  \caption{Turbine and generator rotational speeds, and mechanical and electrical powers for one-mass and two-mass mechanical models}
  \label{fig.results2}
\end{figure}

Both parameters and values of the implemented model are summarised in
Table~\ref{tab.wt}. These parameters are tuned by using a set of
typical wind speed values as an input. With this aim, such wind speed
input has been linearly increased from 5~$m/s$ to 20~$m/s$ in
approximately 150~$s$ time interval, see Figure~\ref{fig.wind1}. This
wind speed variation is not a realistic evolution, but it represents
an educational approach to analyse the evolution of the different
parameters under wind speed variations.  The results for the pitch
angle, tip speed ratio, and power coefficient are summarised in
Figure~\ref{fig.results1} for the one and two-mass mechanical models. As
can be seen, pitch angle results ($\beta$) are very similar for both
mechanical models, see Figure~\ref{fig.pitch1}. However, more
significant differences are obtained for tip speed ratio ($\lambda$)
and power coefficient ($C_{p}$), see Figure~\ref{fig.tsr1} and
~\ref{fig.cp1} respectively. These discrepancies are directly related
to the differences between $\Omega_{WT}$ to be applied on each
mechanical model. To clarify this point, Figure~\ref{fig.results2}
depicts the rotational speeds of the turbine and generator ---Figure
\ref{fig.omega1} and \ref{fig.omega_G1}---, as well as the mechanical
and electrical power evolution ---Figure \ref{fig.pm1} and
\ref{fig.pe1}---. As can be seen, turbine and generator rotational
speeds of both models have different values, more especially below the
reference value $\Omega_{ref}=1.2$~pu. These differences are mainly
due to the inertia constant values of both mechanical models
---parameters given in Table~\ref{tab.wt}---, which imply slight
differences between $\lambda$ and $C_{p}$ as shown in
Figure~\ref{fig.results1}. With regard to $P_{mech}$ and $P_{e}$,
shown in Figure \ref{fig.pm1} and \ref{fig.pe1}, minor differences are
obtained and similar results can be estimated from both mechanical
models.

 \begin{table}[tbp]
	\small\sf\centering
	\caption{\textcolor{black}{Wind turbine model parameters\cite{miller03}.}} 
	\resizebox{.5\linewidth}{!}{%
		\begin{tabular}{ll}
			\toprule
			Parameter & Value \\ 
			\midrule
			$K_{pp}$ &  150\\
			$K_{ip}$ &  25\\
			$K_{pc}$ &  3\\
			$K_{ic}$ & 30\\
			$T_{PI}$ &  0.01~s\\
			$\beta_{max}$ & 27$^{\circ}$\\
			$\beta_{min}$ & 0$^{\circ}$\\
			$d\beta/dt_{max}$ & 10$^{\circ}$/s\\
			$d\beta/dt_{min}$ & -10$^{\circ}$/s\\
			$P_{e,max}$&1~pu\\
			$P_{e,min}$&0.1~pu\\
			$dP_{e}/dt_{max}$ & 0.45 pu/s\\
			$dP_{e}/dt_{min}$ & -0.45 pu/s\\
			$K_{pt}$ &  3\\
			$K_{it}$ &  0.6\\
			$V_{WT}$ & 1 pu\\
			$K_{rotor}$ & 0.00145\\
			$T_{con}$ & 0.02~s\\
			$T_{f}$ & 5~s\\
			$T_{pc}$ & 0.05~s\\ 
			$H_{WT}$ (one-mass) &  5.19~s\\
			$H_{g}$ (two-mass) &  0.90~s\\
			$H_{WT}$ (two-mass) &  4.29~s\\
			$D$ (two-mass) &  1.5\\
			$K$ (two-mass) &  296.7\\
			$\Omega_{0}$ (two-mass) &  1.335\\
			\bottomrule
		\end{tabular}
	}
	\label{tab.wt}
\end{table}



\section{Conclusion}\label{sec.conclusion}

The high integration of wind energy into power systems has promoted
new employment opportunities and a more relevance of such industrial
sector. As a consequence, educational programs also need to evolve and
upgrade in order to cover these sector requirements. Under this
scenario, this paper describes the simulation of a VSWT proposed to
the students of the `Wind energy' subject in the Universidad
Polit\'{e}cnica de Cartagena (Spain). An open-source software
(Scilab-Xcos) is used and successfully assessed in our
practical sessions. Through the different simulation proposals, the
students are able to analyse graphically the influence among the main
parameters of the wind turbines. Two different mechanical models of
the rotor are introduced and simulated, comparing the differences of the results
obtained between both mechanical models. The simulation experimenting
proved to be an interesting alternative to conventional lectures and a
suitable solution for practical sessions under a free open-source
software environment.

\section*{Acknowledgement}
This work was supported by ‘Ministerio de Educación, Cultura y Deporte’ of Spain (FPU16/04282).


\bibliographystyle{SageV}
\bibliography{biblio}

\end{document}